\documentclass[fleqn,twoside]{article}
\usepackage{espcrc2}

\title{O(a) improved QCD: The 3-loop beta-function, and the critical hopping
    parameter\thanks{Presented by H. Panagopoulos}}

\author{A. Bode\address{CSIT, Tallahassee, USA},
        H. Panagopoulos\address{Department of Physics, University of
        Cyprus},
        Y. Proestos$^{b,}$\address{Present address: Department of
        Physics, Ohio State University, USA}}

\begin{document}

\begin{abstract}
We calculate the 3-loop bare $\beta$-function of QCD, formulated on the lattice with
the clover fermionic action. The dependence of our result on the number of colors $N$,
the number of fermionic flavors $N_f$, and the clover parameter $c_{SW}$, is
shown explicitly. A direct outcome of our calculation is the two-loop relation between
the bare coupling constant $g_0$ and the one renormalized in the $\overline{{\rm
MS}}$ scheme. Further, we can immediately derive the three-loop correction to the
relation between the lattice $\Lambda$-parameter and $g_0$, which turns out to be
very pronounced. 

We also calculate the critical value of the hopping parameter, $\kappa_c$, in the 
clover action, to two loops in perturbation theory. This quantity is
an additive renormalization; as such, it exhibits a linear divergence
in the lattice spacing. We compare our results to non perturbative
evaluations of $\kappa_c$ coming from MC simulations. 

\end{abstract}

\maketitle

\section{INTRODUCTION}
\label{introduction}

The clover action for lattice fermions was introduced a number
of years ago~\cite{SW}, as a means of reducing finite lattice spacing
effects. It is widely
used nowadays in Monte Carlo simulations.

To monitor the onset of the continuum limit, tests of scaling must be
performed on measured quantities. In particular, asymptotic scaling is
governed by the bare $\beta$-function:

\begin{equation}
\beta_L(g_0)= -a{dg_0\over da} \mid_{g,\,\mu}, 
\label{lattb}
\end{equation}
($a$ is the lattice spacing, $g$ $(g_0)$ the renormalized (bare)
coupling constant, $\mu$ the renormalization scale). 
For $g_0\rightarrow 0$ one may write $\beta_L$ as:
\begin{equation}
\beta_L(g_0) =
-b_0 \,g^3_0 -b_1 \,g_0^5 - b_2^{L}\,g_0^7 + ...,
\end{equation}

The first two coefficients, $b_0$ and $b_1$, are
universal and well known in $SU(N)$ gauge theory with $N_f$ fermion
species; the 3-loop coefficient $b_2^L$, on the other hand, is 
regularization dependent. In the case at hand, $b_2^L$ is thus
expected to depend not only on $N$ and $N_f$, but also on the
parameter $c_{\rm SW}$ of the clover action (see next Section).

We calculate $b_2^L$ for arbitrary $N$, $N_f$ and
$c_{\rm SW}$. The analogous calculation for pure gauge theory
without fermions~\cite{LWpaper,A-F-P}, as well as for
Wilson fermions~\cite{C-F-P-V-98}, was done a few years ago. We follow
the general setup of those publications.

The $\beta$-function enters directly into the relation defining the
parameter $\Lambda_L$ :
\[
a\Lambda_L = \exp \left( -{1\over 2b_0g_0^2}\right)
(b_0g_0^2)^{-{b_1/2b_0^2}}\cdot \]
\begin{equation}
\qquad\cdot\left[ 1 + q\, g_0^2 + ...\right], \quad
q = (b_1^2-b_0b_2^L)\, / \, 2b_0^3.
\label{asympre}
\end{equation}
The ``correction'' factor $q$ turns out to be very pronounced for
typical values of $c_{\rm SW}$ and $g_0$.

A direct outcome of our calculation is the two-loop relation between
the $\overline{\rm MS}$ coupling $\alpha\equiv g^2/(4\pi)$
and $\alpha_0\equiv g_0^2/(4\pi)$ :
\begin{equation}
\alpha = \alpha_0 + d_1(a\mu)\,\alpha_0^2 +
d_2(a\mu)\,\alpha_0^3+O\left( \alpha_0^4\right),
\label{alpha}\end{equation}
This relation is useful in studies
involving running couplings or renormalized quark masses.

In Sec.~\ref{sec2} we present our results for $b_2^L$,
$d_1(a\mu)$ and $d_2(a\mu)$, as functions of $N$, $N_f$ and
$c_{\rm SW}$. Further technical details and checks of our
calculations are relegated to a longer write-up~\cite{BP}.

We have also calculated the critical value of the hopping
parameter $\kappa_c$, to two
loops, using the clover action. 
Since Wilson fermions break chiral
invariance explicitly, merely setting their bare mass to
zero does not ensure chiral symmetry in the
continuum limit; quantum corrections introduce an additive
renormalization to the fermionic mass, which must then be fine tuned
to a vanishing renormalized value. Consequently, the
hopping parameter $\kappa$
is shifted from its naive value.

The additive mass renormalization is
linearly divergent with the lattice spacing. This adverse 
feature of Wilson fermions
poses an additional problem to a 
perturbative treatment. Indeed, our calculation serves as a
check on the limits of applicability of perturbation theory, by
comparison with non perturbative Monte Carlo results.

In the present work we follow the procedure of
Ref.~\cite{FP}, in which $\kappa_c$ 
was computed using Wilson fermions without
${\cal O}(a)$ improvement. 
In Sec.~\ref{sec3} we present our results on $\kappa_c$, showing
explicitly the dependence on $N$, $N_f$ and $c_{\rm SW}$. Details on our
calculation can be found in our publication~\cite{PP}.

\section{THE $\beta$ FUNCTION}
\label{sec2}

Our starting point is the Wilson formulation of the QCD action on the
lattice, with the addition of the clover~\cite{SW} fermion
term, $S_{\rm SW}$, which reads in standard notation:
\begin{eqnarray}
S_{\rm SW} &=& {i\, a^5\over 4}\,c_{\rm SW}\sum_{x,\,\mu,\,\nu,\,f} \bar{\psi}_{f}(x)
\sigma_{\mu\nu} {\hat F}_{\mu\nu}(x) \psi_f(x), \nonumber \\
{[}{\hat F}_{\mu\nu} &\equiv& {1\over8}\,
(Q_{\mu\nu} - Q_{\nu\mu}), \label{latact}\\
Q_{\mu\nu} &\equiv& U_{\mu,\,\nu} + U_{\nu,\,{-}\mu} +
U_{{-}\mu,\,{-}\nu} + U_{{-}\nu,\,\mu}{]} \nonumber
\end{eqnarray}
Here $U_{\mu,\,\nu}(x)$ is the usual product of link variables
$U_{\mu}(x)$ along the perimeter of a plaquette in the $\mu$-$\nu$
directions, originating at $x$;
$f$ is a flavor index; $\sigma_{\mu\nu} =
(i/2) [\gamma_\mu,\,\gamma_\nu]$.
The value of the parameter $c_{\rm SW}$ can be chosen arbitrarily; it is normally
tuned in a way as to minimize ${\cal O}(a)$ effects.
The lattice $\beta$-function is independent of the
renormalized fermionic masses, which may be set to zero.

We compute the relation between $g_0$ 
and $g$, defined in the $\overline{\rm MS}$
renormalization scheme: 
\begin{equation}
g_0 = Z_g(g_0,\,a\mu) \, g.
\label{grenorm}
\end{equation} 
The one- and two-loop terms of $Z_g^2$ have the form:
$$
Z_g(g_0,\,a\mu)^2  = 1 + L_0(a\mu) \,g_0^2 + L_1(a\mu)\, g_0^4 +
\ldots
$$
\begin{equation}
L_0(x) = 2b_0 \ln x + l_0,\quad L_1(x) = 2b_1 \ln x + l_1.
\label{zgzg}\end{equation}

The constant $l_0$ is related to the ratio 
of the associated $\Lambda$ parameters:
\begin{equation}
l_{0} = 2b_0\ln \left( \Lambda_L/
\Lambda_{\overline{\rm MS}}\right).
\end{equation}
Its value is known (see
e.g. Ref.~\cite{BWW} and references therein) and is presented here
with increased accuracy for the $c_{\rm SW}$-dependent coefficients:
$$l_0 ={1\over 8N} -0.1699559991998031(2)\, N + N_f\, l_{01} \label{l0}$$
\begin{eqnarray}l_{01} &=& 
 0.006696001(5) -  c_{\rm SW}\, 0.00504671402(1) \nonumber\\
&& + {c_{\rm SW}}^2\, 0.02984346720(1)
\end{eqnarray}
The dependence on $c_{\rm SW}$ is quite pronounced,
leading to changes in $\Lambda_L$ of up to a factor of 2. 

The quantity $b_2^L$ can be obtained from $l_0$, $l_1$ :
\begin{equation}
b_2^L= b_2 -b_1l_{0}+ b_0 l_{1}.
\label{b2lrel}
\end{equation}
where $b_2$ is known from the continuum.
Computing $l_1$ amounts to a two-loop calculation of the one-particle irreducible
two-point function of a background gauge field. We find:
\begin{eqnarray}
l_1 &=& - 3/(128 N^2) + N_f\left[\, l_{11}\, /N + l_{12}\,
N\right] \nonumber \\
&& + 0.018127763034(4) - N^2 \, 0.0079101185(2) \nonumber\\
l_{11} &=&-0.0011967(14)+c_{\rm SW}\, 0.0001578(3)\nonumber\\
 &&+{c_{\rm SW}}^2\, 0.0052931(2) +{c_{\rm SW}}^3 \,0.00050624(3)
 \nonumber\\
&&+{c_{\rm SW}}^4\, 0.00008199(1), \nonumber \\
l_{12} &=& 0.0009998(16)+c_{\rm SW} \, 0.0000342(4) \nonumber\\
&&-{c_{\rm SW}}^2\, 0.0048660(6) -{c_{\rm SW}}^3\, 0.00021431(3) \nonumber\\
&&-{c_{\rm SW}}^4\, 0.00004382(1).
\label{l11l12}
\end{eqnarray}

Substitution in Eq.~(\ref{b2lrel}) now yields $b_2^L$, for any
value of $N$, $N_f$, $c_{\rm SW}$.
The correction coefficient $q$ of Eq.~(\ref{asympre}) also follows
immediately; it brings about a substantial correction to asymptotic
scaling, with a pronounced $c_{\rm SW}$ dependence. 
Finally, the coefficients $d_1(a\mu)$ and $d_2(a\mu)$ can be read off 
Eqs.~(\ref{l11l12}), (\ref{alpha}), (\ref{grenorm}), (\ref{zgzg}).

There exist several constraints on the algebraic and numerical values
of individual diagrams.
A particularly strong check is provided by Ref.~\cite{BWW}.
Eq. (5.6) in that reference reads:  
$d_2(N{=}3,\,c_{\rm SW}^{(0)}{=}1) = 1.685(9) - 8.6286(2) \, c_{\rm SW}^{(1)}$. 
For the same quantity, our results lead to: 
$1.6828(8) - 8.62843775(1) \, c_{\rm SW}^{(1)}$. 
Both sets of numbers are clearly in very good agreement.

\section{THE HOPPING PARAMETER}
\label{sec3}

We set out to calculate the hopping parameter, 
\begin{equation}
\kappa\equiv 1\, / \, (2\,m_B\,a + 8\,r),\label{kappa}
\end{equation}
which is an adjustable quantity in numerical simulations; $m_B$ is the
bare fermionic mass, and $r$ 
is the Wilson parameter appearing in the fermionic action,
usually set to 1. The critical value of $\kappa$, at which chiral symmetry
is restored, is thus $1/8r$ classically, but gets shifted by quantum
effects.

For a vanishing renormalized mass we require:
\begin{equation}
m_B = \Sigma^L(0,m_B,g_0)\label{mB}
\end{equation}
where $\Sigma^L(p,m_B,g_0)$ is the truncated,
1PI fermionic two-point function.
We solve this recursive equation for $m_B$ perturbatively. We write:
\begin{equation}
\Sigma^L(0,m_B,g_0) = g_0^2 \, \Sigma^{(1)} + g_0^4 \, \Sigma^{(2)} + \cdots
\end{equation}

Two diagrams contribute to $\Sigma^{(1)}$, and 26 to $\Sigma^{(2)}$.
Certain sets of diagrams must be evaluated together
for infrared convergence.
The dependence on $c_{\rm SW}$ is polynomial.
For $\Sigma^{(1)}$ we find: 
\begin{equation}
\Sigma^{(1)}= (N^2-1) \,/\, N \,
\bigl(-0.16285705871085(1) \label{Sigma1}
\end{equation}
$$\qquad +c_{\rm SW}\, 0.04348303388205(10)$$
$$\qquad +{c_{\rm SW}}^2 \, 0.01809576878142(1)\, \bigr)$$
One- and two-loop results pertaining to $c_{\rm SW} = 0$ are as in
Ref.~\cite{FP}, and can be found with greater accuracy in a subsequent work~\cite{CPR}.

For $c_{\rm SW}\ne 0$, only one-loop results exist so far in the
literature; a recent presentation for $N{=}3,\ c_{\rm SW}{=}1$~\cite{BWW} agrees
with our Eq.~(\ref{Sigma1}):
\[
\begin{array}{rll}
\Sigma^{(1)} \to &-0.2700753495(2),    &{\rm Ref.~\cite{BWW}}\\
                 &-0.2700753494597(5), &{\rm Eq.~(\ref{Sigma1})}.
\end{array}
\]

We now turn to the much more cumbersome
evaluation of $\Sigma^{(2)}$.
Our result is:
\begin{equation}
\begin{array}{l}
\Sigma^{(2)} / (N^2 {-} 1)=\\
\quad\bigl[(-0.017537(3)   +1/N^2\ 0.016567(2) \\ 
\qquad\qquad +N_f/N\ 0.00118618(8) )\\
\quad {+}(0.002601(2)   -1/N^2\ 0.0005597(7) \\
\qquad\qquad -N_f/N\ 0.0005459(2) )\, c_{\rm SW} \\
\quad {+}(-0.0001556(3)  +1/N^2\ 0.0026226(2) \\
\qquad\qquad +N_f/N\ 0.0013652(1) ) \, {c_{\rm SW}}^2\\
\quad {+}(-0.00016315(6) +1/N^2\ 0.00015803(6) \\
\qquad\qquad -N_f/N\ 0.00069225(3) ) \,{c_{\rm SW}}^3\\
\quad {+}(-0.000017219(2)+1/N^2\ 0.000042829(3) \\
\qquad\qquad -N_f/N\ 0.000198100(7) )\,{c_{\rm SW}}^4 \bigr]
\label{Sigma2}
\end{array}
\end{equation}

Several important consistency checks can be performed on the values of
individual diagrams; our results satisfy these checks~\cite{PP}.
Eqs.~(\ref{Sigma1}, \ref{Sigma2}) lead immediately to the critical
mass and hopping parameter, making use of Eqs.~(\ref{mB}, \ref{kappa}).

A number of non-perturbative determinations of $\kappa_c$ 
exist for particular values of 
$g_0$, $c_{\rm SW}$, and $N_f$.
We present these in~\cite{PP}, along with our
results and with
``dressed'' results obtained by a resummation of cactus
diagrams~\cite{cactus}. 
Comparing with the Monte Carlo estimates, dressed results show a
definite improvement over non-dressed values. 


\end{document}